\documentclass{article}
\usepackage{frExample}
\usepackage{apalike}
\usepackage{cite}
\usepackage{amsmath,amssymb,amsfonts}
\usepackage{algorithmic}
\usepackage{graphicx}
\usepackage{textcomp}
\usepackage{xcolor}
\usepackage{array, booktabs}

\usepackage{enumitem}
\usepackage{dsfont}
\usepackage{siunitx}
\usepackage{float}
\usepackage{rotating}
\usepackage{placeins}

\usepackage[hyphens]{url}
\PassOptionsToPackage{hyphens}{url}
\usepackage{hyperref}

\def\BibTeX{{\rm B\kern-.05em{\sc i\kern-.025em b}\kern-.08em
    T\kern-.1667em\lower.7ex\hbox{E}\kern-.125emX}}

\urldef{\vidurl}\url{https://www.youtube.com/playlist?list=PLEkfZ4KJSCcG4yGWD7K5ENGXW5nBPYiF1}
\urldef{\codeurl}\url{https://www.github.com/ribsthakkar/HierarchicalKarting}

\makeatletter
\newcommand\extralabel[2]{{\edef\@currentlabel{\@currentlabel#2}\label{#1}}}
\makeatother
\title{Hierarchical Control for\\Head-to-Head Autonomous Racing}

\author{
Rishabh Saumil Thakkar
\\
Oden Institute for Computational Engineering and Sciences\\
The University of Texas at Austin\\
Austin, TX 78712 \\
\texttt{rishabh.thakkar@utexas.edu} \\
\And
Aryaman Singh Samyal\\
Department of Aerospace Engineering and Engineering Mechanics\\
The University of Texas at Austin\\
Austin, TX 78712 \\
\texttt{aryamansinghsamyal@utexas.edu} \\
\And
David Fridovich-Keil \\
Department of Aerospace Engineering and Engineering Mechanics\\
The University of Texas at Austin\\
Austin, TX 78712 \\
\texttt{dfk@utexas.edu} \\
\And
Zhe Xu \\
School for Engineering of Matter, Transport, and Energy\\
Arizona State University\\
Tempe, AZ 85281 \\
\texttt{xzhe1@asu.edu} \\
\And
Ufuk Topcu \\
Department of Aerospace Engineering and Engineering Mechanics\\
The University of Texas at Austin\\
Austin, TX 78712 \\
\texttt{utopcu@utexas.edu} \\
}

\begin{document}

\maketitle

\begin{abstract}
We develop a hierarchical controller for head-to-head autonomous racing. We first introduce a formulation of a racing game with realistic safety and fairness rules. A high-level planner approximates the original formulation as a discrete game with simplified state, control, and dynamics to easily encode the complex safety and fairness rules and calculates a series of target waypoints. The low-level controller takes the resulting waypoints as a reference trajectory and computes high-resolution control inputs by solving an alternative approximation formulation with simplified objectives and constraints. We consider two approaches for the low-level planner, constructing two hierarchical controllers. One approach uses multi-agent reinforcement learning (MARL), and the other solves a linear-quadratic Nash game (LQNG) to produce control inputs. The controllers are compared against three baselines: an end-to-end MARL controller, a MARL controller tracking a fixed racing line, and an LQNG controller tracking a fixed racing line. Quantitative results show that the proposed hierarchical methods outperform their respective baseline methods in terms of head-to-head race wins and abiding by the rules. The hierarchical controller using MARL for low-level control consistently outperformed all other methods by winning over 90\% of head-to-head races and more consistently adhered to the complex racing rules. Qualitatively, we observe the proposed controllers mimicking actions performed by expert human drivers such as shielding/blocking, overtaking, and long-term planning for delayed advantages. We show that hierarchical planning for game-theoretic reasoning produces competitive behavior even when challenged with complex rules and constraints.
\end{abstract}

\section{Introduction}
Autonomous driving has seen an explosion of research in academia and industry \cite{adlit}. While most of these efforts focus on day-to-day driving, there is growing interest in autonomous racing \cite{litreview}. Many advancements in commercial automobiles have originated from projects invented for use in motorsports such as disc brakes, rear-view mirrors, and sequential gearboxes \cite{racingadvances}. The same principle can motivate the design of self-driving controllers because racing provides a platform to develop these controllers to be highly performant, robust, and safe in challenging scenarios.

Successful human drivers are required to both outperform opponents and adhere to the rules of racing. These objectives are effectively at odds with one another, but the best racers can satisfy both. Prior approaches in autonomous racing usually over-simplify the latter by only considering collision avoidance \cite{Wang2019, Wang2021, Li2021, He2021}. In reality, these racing rules often involve discrete variables and complex nuances \cite{racingrules}. For example, a driver may not change lanes more than a fixed number of times when traveling along a straight section of the track. While it is relatively straightforward to describe this rule in text, it is challenging to encode the rule in a mathematical formulation of the game, which can be solved by existing methods for real-time control. These methods have to compromise by either shortening their planning horizons or simply ignoring these constraints. The resulting behavior is an agent that is not optimal, or an agent that may be quick but is unsafe or unfair.

We develop a hierarchical control scheme that reasons about the optimal long-term plans and closely adheres to the safety and etiquette rules of a racing game.
The contributions include constructing a mathematical formulation of a head-to-head racing game that incorporates the nuanced lane-changing and collision avoidance rules, developing the hierarchical controller consisting of a high-level tactical planner and a low-level path planner, and verifying in a high-fidelity simulation that our method outperforms various baselines representing common prior approaches. By incorporating the nuanced safety and etiquette rules in our proposed formulation, we construct a mixed-integer nonlinear problem due to the discrete nature of those constraints. As a result, we use a hierarchical structure to solve two approximations of the proposed problem yielding the aforementioned planners.

\begin{figure}
  \centering
  \includegraphics[width=0.85\textwidth]{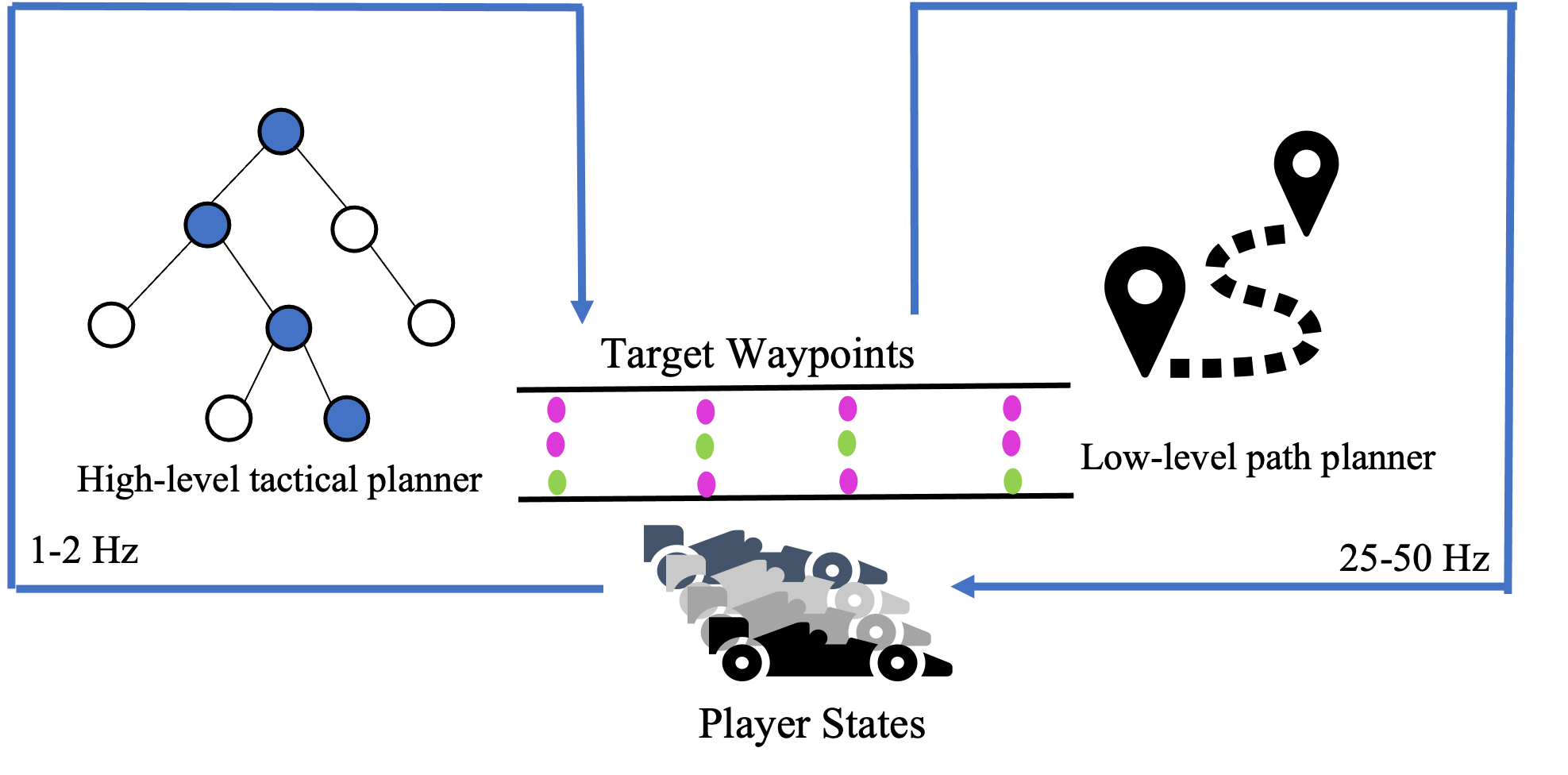}
  \caption{Two-level planning architecture of the proposed racing controller.}
  \label{fig:control_arch}
\end{figure}

The high-level tactical planner constructs a fully-discretized abstraction of the original game allowing us to easily encode the complex rules and plan with a longer time horizon. The solution of the high-level game produces a series of discrete target waypoints. The low-level path planner constructs a simplified version of the original game with an objective to track the high-level planner's target waypoints and a simplified set of the rules because it assumes the tactical planner has already produced a path that accounted for the complex rules. Using our simplified abstractions, we exploit existing methods such as Monte Carlo tree search, linear-quadratic Nash game, and multi-agent reinforcement learning to easily solve the simplified games. The resulting control system runs in real-time and outperforms other traditional control methods in terms of head-to-head performance and a measure of obedience to the safety rules. Figure \ref{fig:control_arch} visualizes the overall control architecture. 

\section{Prior Work}
 There are many levels in the stack of a complete autonomous racing platform that have been studied and are to be studied as discussed in the recent literature review by \cite{litreview}. This paper focuses on the path planning layer. Furthermore, most prior work in autonomous racing related to path planning is focused on single-agent lap time optimization, with fewer and more recent developments in multi-agent racing because multi-agent racing is inherently more complex.

Single-agent racing approaches utilize a mixture of optimization and learning-based methods. One study uses Monte Carlo tree search to estimate where to position the car around various shaped tracks to define an optimal trajectory \cite{Hou2016}. The work in \cite{Vazquez2020} proposes a method that computes an optimal trajectory offline and uses a model predictive control (MPC) algorithm to track the optimized trajectory online. Similarly, the authors of \cite{Stahl2019_2} also perform calculations offline by creating a graph representation of the track to compute a target path and use spline interpolation for online path generation in an environment with static obstacles. In the category of learning-based approaches, online learning to update parameters of an MPC algorithm based on feedback from applying control inputs is developed in \cite{Kabzan2019}. Recent work in \cite{Remonda2021, deBruin2018, weiss2020} develops and compares various deep reinforcement learning methods to find and track optimal trajectories. The results in \cite{ghignone2023tc, evans2023} improve the robustness and generalizability of deep reinforcement learning techniques for autonomous racing. 

State-of-the-art methods in multi-agent racing also use both optimization and learning-based control approaches. Authors of \cite{Li2021} use mixed-integer quadratic programming formulation for head-to-head racing with realistic collision avoidance but concede that this formulation struggles to run in real time. \cite{li2023thesis} improves the previously mentioned method based on integer programming by solving two smaller, simplified problems considering trajectories only overtaking on the left or right of the opponent and picking the optimal solution of the two. Another study proposes a real-time control mechanism for a game with a pair of racing drones \cite{spica2020real}. This work provides an iterative best-response method while solving an MPC problem that approximates a local Nash equilibrium. It is eventually extended to automobile racing \cite{Wang2019} and multi-agent scenarios with more than two racers \cite{Wang2021}. A faster, real-time MPC algorithm to make safe overtakes is developed in \cite{He2021}, but their method does not consider adversarial behavior from the opposing players. \cite{jia2023} discusses a method for drone racing formulated as a dynamic potential game that enables the controller to solve a single optimal control problem rather than solving multiple using iterative best-response. It is worth emphasizing that these approaches do not consider racing rules other than simple collision avoidance. The work in \cite{sonyai} develops an autonomous racing controller using deep reinforcement learning that considers the rules of racing beyond just simple collision avoidance. Their controller outperforms expert humans while also adhering to proper racing etiquette. It is the first study to consider nuanced safety and fairness rules of racing and does so by developing a reward structure that trains a controller to understand when it is responsible for avoiding collisions, and when it can be more aggressive. 

Finally, hierarchical game-theoretic reasoning is a method that has been previously studied in the context of autonomous driving. A hierarchical racing controller was introduced in \cite{LinigerThesis} that constructed a high-level planner with simplified dynamics to sample sequences of constant curvature arcs and a low-level planner that used MPC to track the arc that provided the furthest progress along the track. A two-level planning system is developed in \cite{Fisac2019} to control an autonomous vehicle in an environment with aggressive human drivers. The upper-level system produces a plan to be safe against the uncertainty of the human drivers in the system by using simplified dynamics. The lower-level planner implements the strategy determined by the upper level-planner using precise dynamics.

\section{General Head-to-Head Racing Game Formulation} \label{sec:gen_form}
To motivate the proposed control design, we first present a dynamic game formulation of a head-to-head racing game that includes realistic lane-changing and collision avoidance rules. Table \ref{tab:gen_symbols} lists all of the variables and functions referenced in this formulation as well as all other discussed equations and formulations in the following sections. 

Let there be two players, $i$ and $j$, racing over $T$ steps in $\mathcal{T} = \{1, ..., T\}$. To simplify future notation define $N$ as the set consisting of the two players. There is a track defined by a sequence of $\tau$ checkpoints along the center, $\{c_i\}_{i=1}^{\tau}$, whose indices are in a set $C=\{1,..., \tau\}$. The objective for each player $i$ is to minimize the difference between its own time to reach the final checkpoint and player $j$'s time to reach its final checkpoint, i.e. $c_\tau$. We use $\gamma^i$ to represent the earliest time step when a player $i$ reaches the final checkpoint. In effect, each player targets to reach the finish line before its opponent and with the largest time advantage. The continuous state (such as position, speed, or tire wear) for a player $i$, denoted as $x^i_t \in X \subseteq \mathbb{R}^n$, and control, denoted as $u^i_t \in U \subseteq \mathbb{R}^k$, are governed by known dynamics\footnote{We do not provide an explicit set of dynamics equations nor a list of the state variables because the general formulation is mostly independent of that choice. Any model for vehicle dynamics with any number of variables can be used here as long as there is some notion of position and velocity. We approximate these dynamics using a unicycle model in Section \ref{sec:ll_form:lqng} when modeling one of our low-level planners.} $f^i$. We also introduce a pair of discrete state variables $r^i_t \in C$ and $\gamma^i \in \mathcal{T}$. The index of the latest checkpoint passed by player $i$ at time $t$ is $r^i_t$, and it is computed by the function $p: X\rightarrow C$.  We assume these definitions also exist for player $j$. Using these definitions, we formulate the objective, Equation \eqref{eq:gen_obj}, and core dynamics, Equations \eqref{eq:gen_dyn}-\eqref{eq:gen_goal_time}, of the game from player $i$'s perspective as follows:

\begin{equation} \label{eq:gen_obj}
    \min_{u^i_0, ..., u^i_T} \gamma^i - \gamma^j \\
\end{equation}
\begin{equation*}
    \text{s.t.}
\end{equation*}
\begin{equation} \label{eq:gen_dyn}
    x^k_{t+1} = f(x^k_t, u^k_t), \quad \forall \;\; t \in \mathcal{T}, k \in N
\end{equation}
\begin{equation} \label{eq:gen_idx_map}
    r^k_{t+1} = p(x^k_{t+1}, r^k_t), \quad \forall \;\; t \in \mathcal{T}, k \in N
\end{equation}
\begin{equation} \label{eq:gen_init_idx}
    r^k_{1} = 1, \quad \forall \;\; k \in N
\end{equation}
\begin{equation} \label{eq:gen_reach_goal}
    r^k_{T} = \tau, \quad \forall \;\; k \in N
\end{equation}
\begin{equation} \label{eq:gen_goal_time}
    \gamma^k = \min \{t \, | \, r^k_t = \tau \wedge t \in \mathcal{T} \}, \quad \forall \;\; k \in N
\end{equation}

In addition to the core dynamics of the game, some rules govern the players' states. To ensure that the players stay within the bounds of the track we introduce a function, $q: X \rightarrow \mathbb{R}$, which computes a player's distance to the closest point on the center line. This distance must be limited to the width of the track $w$. Therefore, for all $t \in \mathcal{T}$ and $k \in N$:
\begin{equation} \label{eq:gen_idx_dist}
    q(x^k_{t}) \leq w
\end{equation}

Next, we define the collision avoidance rules of the game. We use an indicator function that evaluates if one player is ``behind" another player. The distance between player $i$'s nose and player $j$, computed by function the $d^i: X \rightarrow \mathbb{R}$, is required to be at least $s_1$ if player $i$ is behind player $j$. Otherwise, the players must only be $s_0$ apart. In other words, the players may only be near each other if they are racing side by side. If one is behind the other, the player who is behind must have a larger safety buffer to avoid rear-end collisions. For all $t \in \mathcal{T}$, this rule is expressed by the following constraints for each of the players:
\begin{equation} \label{eq:gen_coll_avoid}
    d^i(x^j_t) \geq  \begin{cases} s_1 & \mathds{1}_{\text{player} \, i \, \text{behind player}\,j \, \text{at}\, t} \\
    s_0 & \text{otherwise}  \end{cases}
\end{equation}

\begin{equation} \label{eq:gen_coll_avoid2}
    d^j(x^i_t) \geq  \begin{cases} s_1 & \mathds{1}_{\text{player} \, j \, \text{behind player}\,i \,\text{at}\, t} \\
    s_0 & \text{otherwise}  \end{cases}
\end{equation}
Finally, the players are limited in how often they may change lanes depending on the part of the track they occupy. We assume that there are $\lambda \in \mathbb{Z^+}$ lanes across all parts of the track. If the player's location on the track is classified as a curve, there is no limit on lane changing. However, if the player is at a location classified as a ``straight,'' it may not change lanes more than $L$ times for that contiguous section of the track. We define a set $\mathcal{S}$ that contains all possible states where a player is located in a straight section. We also introduce a function $z: X \rightarrow \{1, 2, ..., \lambda\}$ that returns the lane ID based on a player's position on the track. Using these definitions, we introduce a variable $l^k_t$, referred to as the count of recent lane changes, calculated by the following constraint for all $t \in \mathcal{T}$ and $k \in N$:
\begin{equation} \label{eq:gen_lane_var}
    l^k_{t} =  \begin{cases} l^k_{t-1} + 1 & \mathds{1}_{x^k_t \, \in \mathcal{S}} = \mathds{1}_{x^k_{t-1} \in \mathcal{S}}\wedge z(x^k_t) \neq z(x^k_{t-1}) \\
    0 & \text{otherwise}  \end{cases}
\end{equation}
This variable effectively represents a player's count of ``recent'' lane changes over a sequence of states located across a contiguous straight or curved section of the track. However, the variable is only required to be constrained if the player is on a straight section of the track. Therefore, the following constraint must hold for all $t \in \mathcal{T}$ and $k \in N$ and if $x^j_t \, \in \mathcal{S}$:
\begin{equation} \label{eq:gen_lane_lim}
    l^k_{t} \leq  L
\end{equation}

Most prior head-to-head/multi-agent racing formulations \cite{Wang2019, Wang2021, Li2021} do not include the complexities we introduced by defining the constraints in Equations \eqref{eq:gen_coll_avoid}-\eqref{eq:gen_lane_lim}. They usually have a similar form regarding continuous dynamics and discrete checkpoints resembling Equations \eqref{eq:gen_dyn}-\eqref{eq:gen_goal_time}, but their rules only involve staying on track, defined in Equation \eqref{eq:gen_idx_dist}, and collision avoidance with a fixed distance regardless of the players' relative positions. However, in real-life racing, there do exist these complexities both in the form of mutually understood unwritten rules and explicit safety rules \cite{racingrules}. As a result, we account for two of the key rules that ensure the game remains fair and safe:
\begin{enumerate}
    \item There is a greater emphasis on and responsibility of collision avoidance for the player that is following the other per Equations \eqref{eq:gen_coll_avoid}-\eqref{eq:gen_coll_avoid2}.
    \item The player may only switch lanes $L$ times while on a straight section of the track per Equations \eqref{eq:gen_lane_var}-\eqref{eq:gen_lane_lim}.
\end{enumerate}

The first rule ensures that a leading player can choose an action without needing to consider an aggressive move that risks a rear-end collision or side collision while turning from the players that are following. This second rule ensures that the leading player may not engage in aggressive swerving or ``zig-zagging" across the track that would make it impossible for a player that is following the leader to safely challenge for an overtake. While functions may exist to evaluate these spatially and temporally dependent constraints, their discrete nature suggests that they cannot be easily differentiated. Therefore, most state-of-the-art optimization algorithms would not apply or struggle to find a solution for real-time control. 

\begin{table}[H]
\centering
\caption{Symbols used in formulations}
\begin{tabular}{l|l}  \label{tab:gen_symbols}
Symbol & Value\\ 
\hline
$N$ & Set of players in the game \\
$\mathcal{T}$ & Set of steps in the game \\
$\delta$ & Number of steps in the game \\
$\{c_i\}_{i=1}^{\tau}$ & Sequence of checkpoints \\
$C$ & Set of checkpoint indices \\
$x^i_t$     &   Continuous state of player $i$ at time $t$   \\
$u^i_t$      &   Control input of player $i$ at time $t$    \\
$f(x^i_t, u^i_t)$  &  Continuous dynamics of player $i$ \\ 
$r^i_t$      &  Index of last checkpoint passed by player $i$ at time $t$ \\ 
$\gamma^i$     &  Time when player $i$ passed last checkpoint \\     
$p(x^i_{t+1}, r^i_t)$  &  Function computing index of last track checkpoint passed  \\     
$q(x^i_{t+1})$   &  Function computing minimum distance to last track checkpoint passed \\
$w$      &  track width \\     
$d^i(x^j_t)$ & Function computing distance between player i's nose and player j \\
$s_0$ & Minimum distance safety margin if the player is not directly behind another \\
$s_1$ & Minimum distance safety margin if the player is directly behind another  \\
$\lambda$ & Number of lanes in the track \\
$l^i_t$ & Integer state variable indicating player $i$'s recent lane changes at time $t$ \\
$y(x^i_t)$ & Function evaluating if player i is on a straight or curve \\
$z(x^i_t)$ & Function computing the lane of the track player occupies\\
$L$ & Upper bound on the number of recent lane changes a player is allowed \\
$\hat{\mathcal{T}}$ & Set of steps in the low-level game \\
$\hat{\delta}$ & Shortened horizon \\
$\alpha$      &   Weight parameter in objective emphasizing the importance of hitting trajectory waypoints    \\
$\psi^i_c$      &   Waypoint for player $i$ to target when passing checkpoint index $c$ \\
$\eta^i_c$  &  Distance of player $i$'s closest approach to the waypoint $\psi^i_c$ \\ 
$h(x^i_{t}, \psi^i_c)$  &  Function distance of player $i$'s state to waypoint $\psi^i_c$ \\  
\end{tabular}
\end{table}
\section{Hierarchical Control Design}
\begin{figure*}
  \centering
  \includegraphics[width=\textwidth]{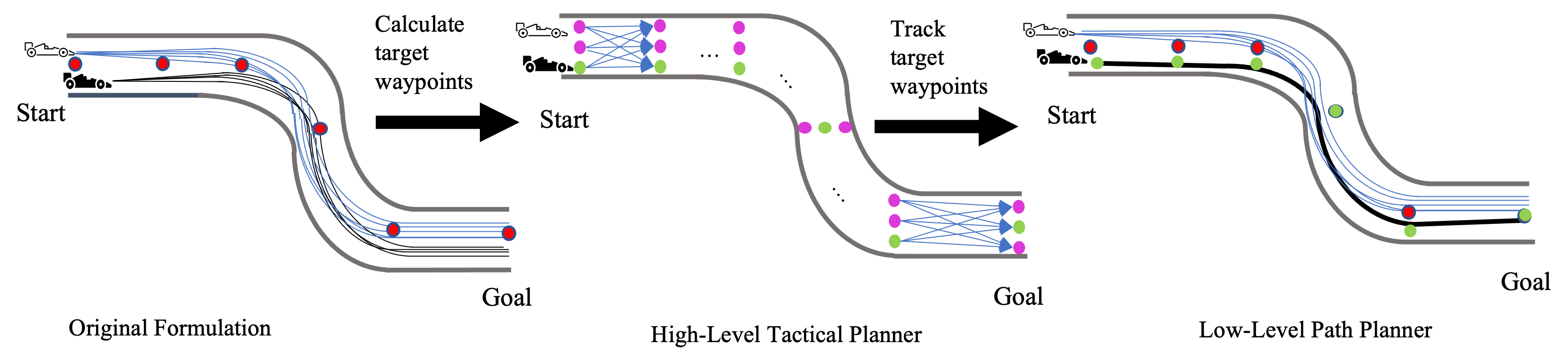}
  \caption{The uncountably infinite trajectories of the general game (left) discretized by the high-level planner (middle). The sequence of target waypoints calculated by the high-level planner (in green) is tracked by the low-level planner (right) and converges to a continuous trajectory (in black).}
  \label{fig:overall_control}
\end{figure*}
Traditional optimization-based control methods cannot easily be utilized for the general head-to-head racing game formulated with realistic safety and fairness rules. The rules involve nonlinear constraints over both continuous and discrete variables, and a mixed-integer nonlinear programming algorithm would be unlikely to run at rates of \SI{25}- \SI{50}{\hertz} for precise control. This inherent challenge encourages utilizing black-box methods such as deep reinforcement learning or trying to solve the game on a short horizon. However, directly applying those methods may result in behavior that does not reliably meet the constraints or is not strategically optimal in the long termWe propose a hierarchical control design involving two parts that work to ensure all of the rules are followed while approximating long-term optimal choices. The high-level planner transforms the general formulation into a game with discrete states and actions where all of the discrete rules are naturally encoded. The solution provided by the high-level planner is a series of discrete states (i.e. waypoints) for each player, which satisfies all of the rules. Then, the low-level planner solves a simplified version of the racing game with an objective putting greater emphasis on tracking the high-level planner's waypoints and a smaller emphasis on the original game-theoretic objective and a simplified version of the rules. Therefore, this simplified formulation can be solved by an optimization method in real time or be reliably trained in a neural network when using a learning-based method. 

\subsection{High-Level Tactical Planner}
The high-level tactical planner constructs a turn-based, discrete game that approximates the general game formulation presented in the previous section. To achieve this, we first transform the continuous state of the original formulation into a discrete representation. Then, we discuss how players transition between states and the dynamics of the discrete game. Finally, we describe the objective of the game and how it is solved, including a visualization of a sample trajectory.

\subsubsection{State Space Model} \label{sec:hl_disc}
 As mentioned in the previous section, we do not assume the use of any specific dynamics model in the general formulation. However, we only assume that there is a notion of velocity and position in a player's continuous state. In the high-level planning model, the velocity and other continuous components of a player's state (except for position) are represented as discrete ranges (e.g., velocity between \SI{2}{\meter\per\second} and \SI{4}{\meter\per\second} or tire wear between 10\% and 15\%). The player's position is represented using a pair of discrete variables that are computed in the constraints of the original formulation: lane ID and last passed checkpoint index, which are computed in the constraints of the original formulation (Equations \eqref{eq:gen_idx_map} and \eqref{eq:gen_lane_var}). However, we only include the lane ID as part of the state because the last passed checkpoint index is used differently as discussed in the next paragraph.
 
\begin{figure}
  \centering
  \includegraphics[width=0.6\textwidth]{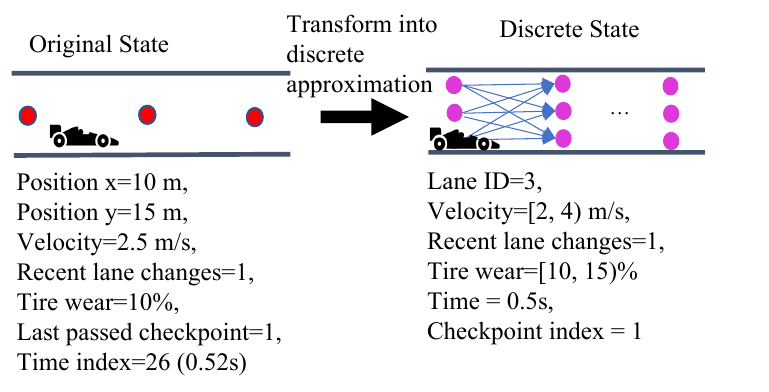}
  \caption{An example of a player's state in the original game (left) with a mix of discrete and continuous elements converted into our high-level tactical planner state approximation (right). The position is converted into a lane ID and checkpoint index. Velocity and tire wear are projected into ranges of some fixed size. The time step is reduced to a finite precision time state in the discrete game; in this example, it is tenths of a second. The recent lane changes state variable remains unchanged because it is inherently discrete.}
  \label{fig:discrete_abs}
\end{figure}

The time index of a player's state in the original formulation is rounded to some finite precision and included as part of the state in the discrete representation. Instead of using the time index to index each player's state, we use the last passed checkpoint index to index the steps of the discrete game. This transformation reduces the size of the state space, allowing us to reason about longer time horizons at the expense of not considering all feasible trajectories. For example, considering a time horizon of \SI{10}{\second} at an average speed of \SI{10}{\meter\per\second} allows the model to plan \SI{100}{\meter} ahead. On the other hand, considering a checkpoint horizon of 8 checkpoints, each \SI{15}{\meter} apart, allows us to look \SI{120}{\meter} ahead. Figure \ref{fig:discrete_abs} visualizes how an example state in the original formulation is transformed into the state space abstraction described in this section.

\subsubsection{Dynamics Model}
Because the index of the discrete game is a track checkpoint, the player's actions are defined by pairs of target lane and the target velocity range for the next checkpoint. Given the location of the target lane at the next checkpoint, the average of the target velocity range, and the state at the current checkpoint, we apply a simplified approximation of the dynamics to compute the state update from one checkpoint to the next. The fidelity of these dynamics is also agnostic to this planning model as long as we can estimate an update for the time it would take to move from one checkpoint to the next that satisfies the boundary conditions. If the boundary conditions are not satisfiable, the choice is ruled out. In our implementation of the high-level planning model, we use simple one-dimensional equations of motion and the provided boundary conditions of the current state and target state to estimate updates to the time state \cite{galilei}. The state variable tracking number of recent lane changes is easily updated by evaluating if the target lane is different from the current lane and if the track type remains the same across the checkpoints, similar to the logic in Equation \eqref{eq:gen_lane_var}.

This action space enables us to easily evaluate or prevent actions where the rules of the game would be broken. By limiting choices to fixed locations at the checkpoints via the lane ID, we ensure that the players always remain on track satisfying the constraint in Equation \eqref{eq:gen_idx_dist}. Moreover, we can dismiss player actions that would violate the limit on the number of lane changes by simply checking whether choosing a lane would exceed their limits or checking if the location is a curve or straight, satisfying the constraint in Equation \eqref{eq:gen_lane_lim}.  Finally, we can dismiss actions that could cause collisions by estimating that if two players reach the same lane at a checkpoint and have a small difference in their time states, there is a high risk of collision, satisfying the constraint in Equation \eqref{eq:gen_coll_avoid}.

The game is played with all players starting at an initial checkpoint, and it progresses by resolving all players' choices one checkpoint at a time. The order in which the players take their actions is determined by the player with the smallest time state at each checkpoint. A lower time state value implies that a player was at the given checkpoint before other players with a larger time state, so it would have made its choice at that location before the others. This ordering also implies that players who arrive at a checkpoint after preceding players observe the actions of those preceding players and use that information in their own strategic choices. Most importantly, because the ordering forces the following players to choose last, we also capture the rule that the following players (i.e. those that are ``behind'' others) are responsible for collision avoidance after observing the leading players' actions. 

\subsubsection{Objective and Solution}
The objective of the discrete game is to minimize the difference between one's own time state at the final checkpoint and the time state of the other player, just like the objective of the original formulation in Equation \eqref{eq:gen_obj}. Although the discrete game is much simpler than the original formulation, the state space still grows exponentially as the number of possible actions and checkpoints increases. To address this, we repeatedly solve smaller versions of the game at \SI{1}{\hertz} using a fixed checkpoint horizon. However, our choice of the horizon extends much further into the future (which we described earlier in Section \ref{sec:hl_disc}) than an MPC-based continuous state/action space controller can handle in real-time \cite{Wang2019}. To solve the discrete game at a frequency of \SI{1}{\hertz}, we use the Monte Carlo tree search (MCTS) algorithm \cite{mcts}. The solution from applying MCTS is a series of waypoints in the form of target lane IDs (which can be mapped back to positions on track given the checkpoint indices) and the target velocities at each of the checkpoints for the ego player and estimates of the best response lanes and velocities for the adversarial player. 

\begin{figure}
  \centering
  \includegraphics[width=0.95\textwidth]{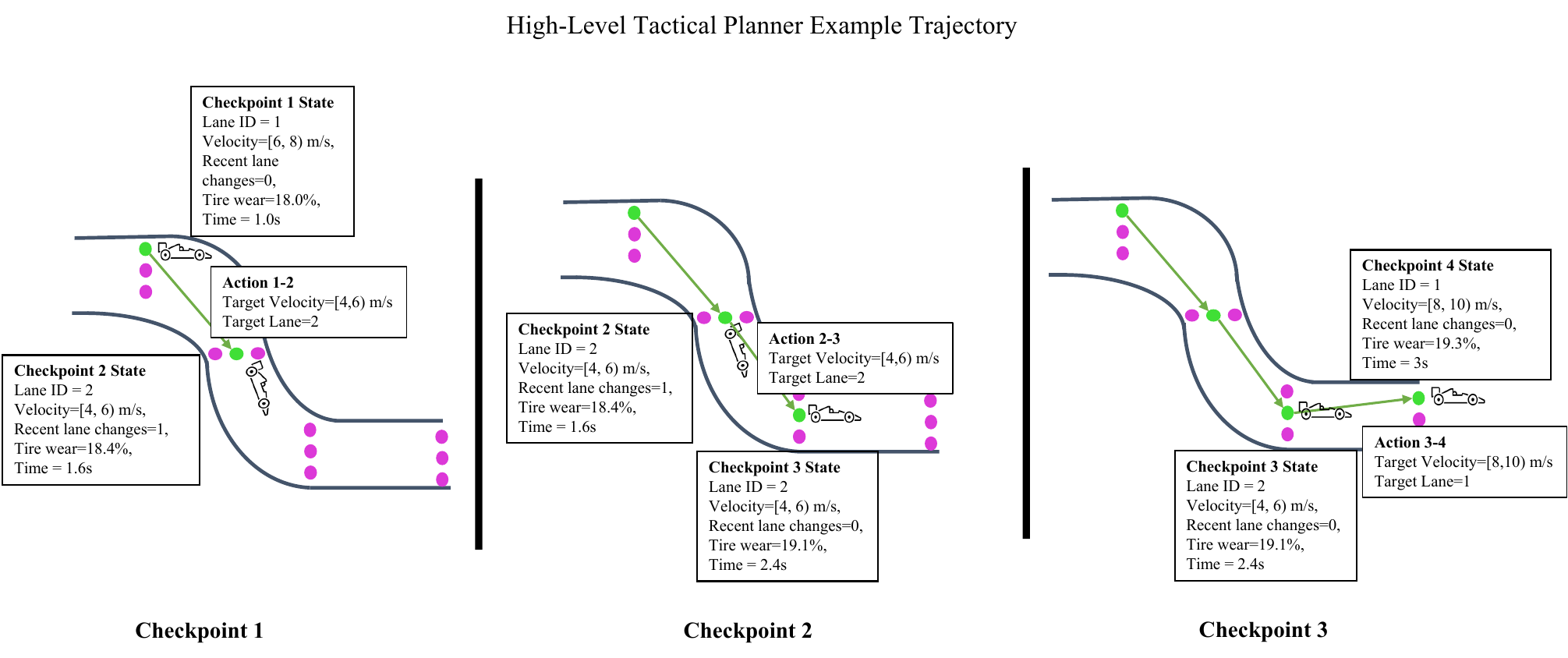}
  \caption{Three steps of an example trajectory for an individual player in the high-level tactical planning model. Components of the state space are discretized or rounded to finite precision. Players choose actions at each checkpoint. Using this model allows us to rapidly simulate long-term plans and evaluate strategies using MCTS.}
  \label{fig:hl_viz}
\end{figure}

Figure \ref{fig:hl_viz} shows an example of a sample trajectory in high-level planner abstraction of a single player. At each checkpoint, the player is located at one of the three lane choices, and the remaining components of its state are expressed with finite precision or as a range of values. Between each checkpoint, the action is labeled by the target velocity range and lane ID of the upcoming checkpoint. The resulting state from applying each of the chosen actions is evaluated using a simplified approximation of the dynamics. The trajectory is simply a path across a directed acyclic graph of lanes and checkpoints, explicitly drawn in the center of Figure \ref{fig:overall_control}. This structure, when all players are included, yields a model that is naturally conducive to utilizing fast tree-search algorithms such as MCTS. 

The high-level abstraction encodes the essence of real-life racing by exploiting the natural discretization that exists in the rules and strategy. For example, the lane-changing rule limits how many times drivers may change lanes on a straight section of the track. This rule relates to the intuition that optimal behavior in racing does not involve frequent changes in direction. Frequent changes in direction impact the weight transfer of the vehicle, which subsequently impacts the grip and stability. If the vehicle is unstable, it is obvious that it is at a higher risk of collision. Furthermore, because drivers do not frequently change directions, they also do not constantly need to make strategic decisions. Instead, they only make strategic choices at certain places on the track, such as the checkpoints, and the remaining time is spent finding the quickest path between the checkpoints. The hierarchical structure is based on this understanding. The solution of the high-level tactical planner is a series of waypoints denoted by target lanes and velocities at each of the checkpoints. It is the job of the low-level path planner to find a trajectory to travel between the planned waypoints.

\subsection{Low-Level Path Planner}
The low-level path planner is responsible for producing the control inputs to follow the target waypoints, and it must operate in real time. We begin by formulating a reduced version of the original game that involves tracking the target waypoints. We then discuss two existing methods to solve the low-level formulation: multi-agent reinforcement learning and a linear-quadratic Nash game approximation.

\subsubsection{Low-Level Formulation} \label{sec:ll_form}
The low-level game is played over a shorter horizon compared to the original game of just $\delta$ steps in $\hat{\mathcal{T}} = \{1, ..., \delta\}$. We assume that the low-level planner for player $i$ has received $k$ waypoints, $\psi^i_{r^i_{1}}, ..., \, \psi^i_{r^i_{1} + k}$, from the high-level planner, and player $i$'s last passed checkpoint $r^i_*$. 
 
The low-level objective involves two components. The first is to maximize the difference between its own checkpoint index and the opponents' checkpoint indices at the end of $\delta$ steps. The second is to minimize the tracking error, $\eta^i_y$, of every passed waypoint $\psi^i_{r^i_{1}+y}$. The former component influences the player to pass as many checkpoints as possible, which suggests reaching $c_\tau$ as quickly as possible. The latter influences the player to be close to the high-level waypoints when passing each of the checkpoints. The objective also includes some multiplier $\alpha$ that balances the emphasis of the two parts. Player $i$'s objective is written as follows:

\begin{equation} \label{eq:ll_obj}
    \min_{u^i_{1}, ..., u^i_{\delta}} (r^j_{\delta} - r^i_{\delta}) + \alpha \sum_{c={r^i_{1}}}^{{r^i_{1}}+k} \eta^i_c
\end{equation}

The players' continuous state dynamics, calculations for each checkpoint, and constraints on staying within track bounds are effectively the same as the original formulation. They are expressed as follows: 

\begin{equation} \label{eq:ll_dyn}
    x^k_{t+1} = f(x^k_{t}, u^k_t), \quad \forall \;\; t \in \hat{\mathcal{T}}, k \in N
\end{equation}
\begin{equation} \label{eq::ll_pos}
    r^k_{t+1} = p(x^k_{t+1}, r^k_t), \quad \forall \;\; t \in \hat{\mathcal{T}}, k \in N
\end{equation}
\begin{equation} \label{eq::ll_pos_init}
    r^k_{1} = r^j_*, \quad \forall\;\; j \in N
\end{equation}
\begin{equation} \label{eq:ll_pos_dist}
    q(x^k_{t}) \leq w, \quad \forall \;\; t \in \hat{\mathcal{T}}, k \in N
\end{equation}

The collision avoidance rules are simplified to just maintaining a minimum distance $s_0$ between the players as we assume that the high-level planner would have already considered the nuances of rear-end collision avoidance responsibilities in Equations \eqref{eq:gen_coll_avoid}-\eqref{eq:gen_coll_avoid2}. As a result, we require the following constraints to hold for all $t \in \mathcal{T}$:
\begin{equation} \label{eq:ll_coll_avoid}
    d^i(x^j_{t}) \geq s_0
\end{equation}
\begin{equation} \label{eq:ll_coll_avoid2}
    d^j(x^i_{t}) \geq s_0
\end{equation}
Finally, we define the dynamics of the waypoint error, $\eta^i_y$, introduced in the objective. It is equivalent to the accumulated tracking error of each target waypoint that player $i$ has passed using a function $h: X\times X \rightarrow \mathbb{R}$ that measures the distance. If a player has not passed a waypoint, then the variable indexed by that waypoint is set to 0. The variable's dynamics are expressed by the following constraint:

\begin{equation} \label{eq:ll_wp_err}
    \eta^i_y = \begin{cases} \sum_{t}^{\Hat{\mathcal{T}}} h(x^i_t, \psi^i_{c})  & \text{if } \exists \; r^i_t \geq y \\
    0 & \text{otherwise}
    \end{cases} \qquad \forall \; y \in \{r^i_{1}, ..., r^i_{1} + k\}
\end{equation}

This simplified formulation is similar to the general formulation in Section \ref{sec:gen_form}, but it has a few key differences. In addition to capturing the goal of finishing ahead of one's opponent, the objective, Equation \eqref{eq:ll_obj}, also includes the notion of minimizing the distance to and passing as many of the high-level planner's waypoints within the time horizon. Second, the complex constraints, Equations \ref{eq:gen_coll_avoid}-\eqref{eq:gen_lane_lim}, introduced by the fairness and safety rules in the general formulation are simplified or dropped. Nevertheless, we still account for the uncertainty of opposing players' actions by maintaining the basic collision avoidance rule in the low-level formulation per Equations \eqref{eq:ll_coll_avoid}-\eqref{eq:ll_coll_avoid2}. 

It is true that because the high-level planner runs at a different frequency and independently of the low-level planner, there is a possibility that the high-level plan calculated based on stale state information or prior to unforeseen deviations from the plan might suggest players to change lanes more frequently than permitted or cause a collision. However, we assume that by generally passing the target waypoints within reasonable error, the low-level path planner's trajectory stays within the subset of trajectories encapsulated by the strategy calculated high-level planner. Under this assumption, we argue that the low-level planner is still generally faithful to the original formulation despite not explicitly considering the original objective and complex rules. The low-level objective influences the agent to be ahead of its opponents and stay close to the high-level plan to ensure the original rules are followed as consistently as possible.  Our assumption is validated in the experiments by analyzing the relationship between a metric that measures the average distance to each of the target waypoints and the number of wins and rule violations for our methods and one of the baselines.

The center and right of Figure \ref{fig:overall_control} show how the waypoints from the high-level planner (in green) are approximately tracked by the low-level planner producing a continuous trajectory (in black). We consider two methods to solve the low-level formulation. The first method develops a reward structure to represent this simplified formulation for a multi-agent reinforcement learning (MARL) controller. The second method further simplifies the low-level formulation into a linear-quadratic Nash game (LQNG) to compute the control inputs. 

\subsubsection{Multi-Agent Reinforcement Learning Controller}
Designing the MARL controller primarily involves shaping a reward structure that models the low-level formulation. We briefly discuss how the rewards model the low-level formulation below, but additional details, such as the mathematical formulation of the rewards, network structure, and training graphs, are presented in the Appendix. 

The RL agent is rewarded for the following behaviors that would improve the objective function \eqref{eq:ll_obj}:
\begin{itemize}
    \item Passing a checkpoint with an additional reward for being closer to the target lane and velocity.
    \item Minimizing the time between passing two checkpoints.
    \item Passing as many checkpoints in the limited time.
\end{itemize}
On the other hand, the agent is penalized for actions that would violate the constraints:
\begin{itemize}
    \item Swerving too frequently on straights \eqref{eq:gen_lane_lim}.
    \item Going off track or hitting a wall \eqref{eq:ll_pos_dist}.
    \item Colliding with other players \eqref{eq:ll_coll_avoid} with additional penalty if the agent is responsible for avoidance \eqref{eq:gen_coll_avoid}. 
\end{itemize}

The rewards capture our low-level formulation objective \eqref{eq:ll_obj} to pass as many checkpoints as possible while minimizing the waypoint error calculated by \eqref{eq:ll_wp_err}, i.e., closely hitting the lane and velocity targets. The penalties capture the on-track, Equation \eqref{eq:ll_pos_dist}, and collision avoidance, Equation  \eqref{eq:ll_coll_avoid}, constraints. However, the penalties also reintroduce the safety and fairness rules, Equations \eqref{eq:gen_coll_avoid}-\eqref{eq:gen_lane_lim}, from the general game formulation that were simplified away from the low-level formulation. Because we assume, as discussed at the end of Section \ref{sec:ll_form}, the rules are generally followed by satisfying the objective of reaching the high-level planner's waypoints, their penalties have weights set much lower than other components of the reward structure. However, we still choose to incorporate the penalties for breaking the original form of the rules to robustify against the possibility that the ego player might be forced to deviate far away from the high-level plan. 

Lastly, the agents' observations include perfect state information of all players, their own target waypoints, and local observations consisting of 9 LIDAR rays spaced over a 180\textdegree{} field of view centered in the direction that the player is facing, which are used to determine the distance to the walls/limits of the track. Note that the observations do not include the opponent's estimated target waypoints, which are also computed during the MCTS calculation by the high-level tactical planner. As a result, the controller's anticipation of the opponent's behavior is fully dependent on the types of maneuvers it might have learned during training using knowledge of the opponent's relative position.

\subsubsection{Linear-Quadratic Nash Game Controller} \label{sec:ll_form:lqng}
Our second low-level approach solves an LQNG using the coupled Riccati equations \cite{basar}. This method involves further simplifying the low-level formulation into a structure with a quadratic objective and linear dynamics. The continuous state is simplified to just four variables: $x$ position, $y$ position, $v$ velocity, and $\theta$ heading. The control inputs $u^i_t$ are also explicitly broken into acceleration, $a^i_t$, and yaw-rate, $e^i_t$. The planning horizon is reduced to $\Bar{\delta}$ where $\Bar{\delta} \ll \delta < T$. To construct our quadratic objective for player $i$, we break it into three components. The first is to minimize the distance to the upcoming target waypoint from the high-level planner $\Bar{\psi}^i$ calculated by the following equation:
\begin{equation} \label{eq:lqng_obj1}
\upsilon^i(\rho_1, \rho_2,\rho_3) =  \sum_{t = 1}^{\Bar{\delta}} (\rho_1((x^i_{t} - \Bar{\psi}^i_x)^2 + (y^i_{t} - \Bar{\psi}^i_y)^2)  + \rho_2 (v^i_{t} - \Bar{\psi}^i_v)^2 
 + \rho_3 (\theta^i_{t} - \Bar{\psi}^i_\theta)^2)
\end{equation}

Because we cannot know how the opponent will behave over the planning horizon, we use the best response computed by MCTS in the high-level planner to estimate the opponent's target waypoints. Using those waypoints, the second component of the objective is to maximize the opponent's distance from the location of those estimated target waypoints $\Bar{\psi^j}$ calculated by the following equation:
\begin{equation} \label{eq:lqng_obj2}
    \phi^i(\Bar{\psi}^j, \rho) = \sum_{t = 1}^{\Bar{\delta}} \rho((x^j_{t} - \Bar{\psi}^j_x)^2 + (y^j_{t} - \Bar{\psi}^j_y)^2)
\end{equation}

We drop all of the constraints except collision avoidance, and it is incorporated as the third component and penalty term in the objective where the distance to each opponent should be maximized. This term is calculated by the following equation:
\begin{equation} \label{eq:lqng_obj3}
    \chi^i(x^j_t, y^j_t, \rho) = \sum_{t = 1}^{\Bar{\delta}} \rho((x^j_{t} - x^i_{t})^2 + (y^j_{t} - y^i_{t})^2)
\end{equation}

The final quadratic objective aggregates the components \eqref{eq:lqng_obj1}-\eqref{eq:lqng_obj3} using weight multipliers ($\rho_i$) to place varying emphasis on the components as follows:

\begin{equation} \label{eq:lqng_obj}
    \min_{a^i_{1}, e^i_{1}, ..., a^i_{\Bar{\delta}}, e^i_{\Bar{\delta}}}
    \upsilon^i(\rho_1, \rho_2,\rho_3)
    - (\phi^i(\Bar{\psi}^j, \rho_4) - \chi^i(x^j_t, y^j_t, \rho_5))
\end{equation}

Finally, the linear dynamics are time-invariant and apply to all players $k \in N$:

\begin{equation} \label{eq:lqng_dyn}
\begin{bmatrix}
x^k_{t+1} \\
y^k_{t+1} \\
v^k_{t+1} \\
\theta^k_{t+1} \\
\end{bmatrix} = 
\begin{bmatrix} 
	1 & 0 & \cos(\theta^k_{t_0})\Delta t & -v^k_{t_0}\sin(\theta^k_{t_0})\Delta t\\
	0 & 1 & \sin(\theta^k_{t_0})\Delta t & v^k_{t_0}\cos(\theta^k_{t_0})\Delta t\\
	0 & 0 & 1 & 0\\
	0 & 0 & 0 & 1\\
	\end{bmatrix}
\begin{bmatrix}
x^k_{t} \\
y^k_{t} \\
v^k_{t} \\
\theta^k_{t} \\
\end{bmatrix} +
\begin{bmatrix} 
	0 & 0 \\
	0 & 0 \\
	\Delta t & 0 \\
	0 & \Delta t \\
	\end{bmatrix}
	\begin{bmatrix} 
	a^k_t  \\
	e^k_t \\
	\end{bmatrix}
\end{equation}

The linear dynamics, in Equation \eqref{eq:lqng_dyn}, are centered around the initial state of the players and are time-invariant. This simplification enables us to easily and rapidly apply the coupled Riccati equations \cite{basar} to compute the control inputs given that the controller must run at \SI{50}{\hertz}. However, an important shortcoming is that the position state variables, $x^k_t$ and $y^k_t$, are only dependent on the acceleration/braking inputs and the initial linearization and do not have a direct dependence on the steering inputs, i.e. the yaw rate. The only incentive for the controller to change the steering input is to match the target steering in the first component of the objective in Equation \eqref{eq:lqng_obj1}. As a result, the drawback of this simplification is that the linearization is only reliable for a small time horizon, so the control inputs are short-sighted. Although we reconstruct and solve the LQNG low-level formulation at \SI{50}{\hertz} to stay up to date with the opponent's state, we are still forced to use more conservative and defensive values for the weights in the model for the collision avoidance objective. We must do so because our estimates of the opponent’s future states are subject to uncertainty in the opponent's behavior. There is no guarantee that the opponent will follow the best response high-level waypoints.

\section{Experiments}
The high-level planner is paired with each of the two low-level planners discussed. We refer to our two hierarchical design variants as MCTS-RL and MCTS-LQNG. 
\subsection{Baseline Controllers}
To measure the importance of our design innovations, we also consider three baseline controllers to resemble the other methods developed in prior works.  

\subsubsection{End-to-End Multi-Agent Reinforcement Learning}
The end-to-end MARL controller referred to as ``E2E," represents the pure learning-based methods such as that of \cite{sonyai}. This controller has a similar reward/penalty structure as our low-level controller, but its observation structure is slightly different. Instead of observing the sequence of upcoming states as calculated by a high-level planner, E2E only receives the subsequence of locations from $\{c_i\}_{i=1}^{\tau}$ that denote the center of the track near the agent. As a result, it is fully up to its neural networks to learn how to plan strategic and safe moves. 

\subsubsection{Fixed Trajectory Linear-Quadratic Nash Game}
The fixed trajectory LQNG controller referred to as ``Fixed-LQNG," uses the same LQNG low-level planner as our hierarchical variant, but it instead tracks a fixed trajectory around the track. This fixed trajectory is a racing line that is computed offline for a specific track using its geometry and parameters of the vehicle as seen in prior works \cite{Vazquez2020, Stahl2019_2}. However, online tracking involves game-theoretic reasoning rather than single-agent optimal control in the prior works.

\subsubsection{Fixed Trajectory Multi-Agent Reinforcement Learning}
The fixed trajectory MARL controller, referred to as ``Fixed-RL," is a learning-based counterpart to Fixed-LQNG. Control inputs are computed using a deep RL policy trained to track precomputed checkpoints that are fixed before the race.  

\subsection{Experimental Setup}
Next, we discuss the experimental setup in the context of the parameters of our simulation environment and the experiments we conduct.
\subsubsection{System Parameters}
Our controllers are implemented\footnote{\codeurl} in the Unity Game Engine. Screenshots of the simulation environment are shown in Figure \ref{fig:experiment_tracks}. We extend the Karting Microgame template \cite{microkarting} provided by Unity. The kart physics implementation uses the kinematic bicycle model \cite{Rajamani2011vehicle}. The kart physics is extended to include cornering limitations and tire wear percentage. Tire wear is modeled as an exponential decay curve that is a function of the accumulated angular velocity endured by the kart. This model captures the concept of losing grip as the tire is subjected to increased lateral loads. We model limits on the lateral acceleration and cornering by limiting the velocity depending on the steering input, tire wear percentage, and turning radius of the vehicle. In effect, we are artificially imposing traction control in the model. As a result, although not explicitly modeled in the formulations, our controllers are required to work around the limitations caused by this mechanic to add another level of realism. We also add multi-agent support is also added to the provided template in to race the various autonomous controllers against each other or human-controlled agents. The high-level planners run at \SI{1}{\hertz}, and low-level planners run at \SI{50}{\hertz} (which is also the limiting rate of the simulation engine). Specifically, $\Bar{\delta}$ is set to \SI{0.06}{\second} for the LQNG planner. The implementation of the learning-based agents utilizes a library called Unity ML-Agents \cite{mlagents}. All of the learning-based control agents are trained using proximal policy optimization and self-play implementations from the library. They are also only trained on two sizes of oval-shaped tracks with the same number of training steps. More details are provided in the Appendix.  

The dynamical parameters such as top speed, acceleration, lateral grip, etc., of each player's vehicle are identical \footnote{The kart's top speed is \SI{15}{\meter\per\second}. It can accelerate up to \SI{4}{\meter\per\square\second} and brake up to \SI{6}{\meter\per\square\second}. It can sustain lateral forces up to 1.5Gs.} The players start every race at the same initial checkpoint. The only difference in their initial states is the lane in which they start. To maintain fairness with respect to starting closer to the optimal racing line, we alternate the starting lanes between each race for the players.

\subsubsection{Setup}
\begin{figure*}
  \centering
  \includegraphics[width=0.98\textwidth]{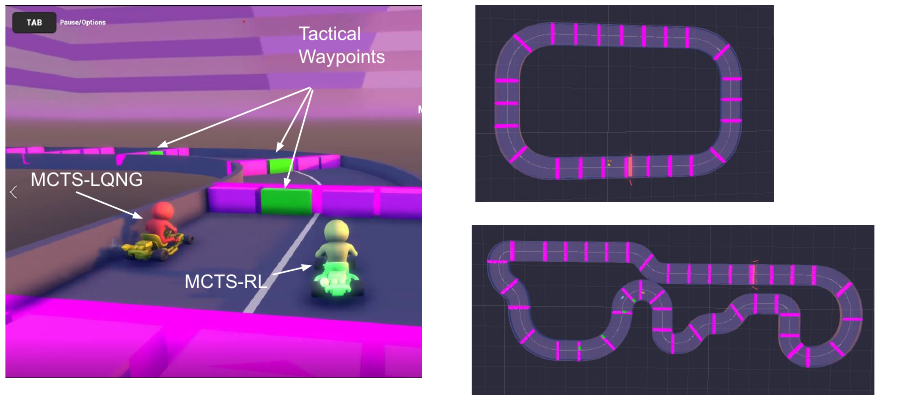}
  \caption{Kart racing environment from a racer's perspective (left), a bird's eye view of the oval track (right-top), and the complex track (right-bottom) in the Unity environment. The purple boxes visualize the lanes across checkpoints along the track, and the highlighted green boxes show planned waypoints determined by MCTS-RL's high-level tactical planner.}
  \label{fig:experiment_tracks}
\end{figure*}

Our experiments include head-to-head racing on a basic oval track (on which the learning-based agents were trained) and a more complex track shown in Figure \ref{fig:experiment_tracks}. Specifically, the complex track involves challenging track geometry with turns whose radii change along the curve, tight U-turns, and turns in both directions. To be successful, the optimal racing strategy requires some understanding of the shape of the track along a sequence of multiple turns. 

The experiments primarily seek to identify the importance of hierarchical game-theoretic reasoning and the effectiveness of the proposed high-level tactical planning model in racing. To achieve this goal and evaluate overall performance, we collect the following data from our experiments: the number of wins against each opponent type, the number of collisions-at-fault, the number of illegal lane changes, and an aggregate safety score (which is the sum of the prior two metrics). In addition, we introduce a pair of metrics, \textit{average target lane distance} denoted as $\Delta_r$ and \textit{average target velocity difference} denoted as $\Delta_v$, for the MCTS-RL, MCTS-LQNG, and E2E agents to evaluate the effectiveness of the high-level tactical planner. Average target lane distance is the average distance between the agent's target lane as provided by the high-level tactical planner and the actual point at which the agent passes a checkpoint in a given race. Similarly, the average target velocity difference is the average difference between an agent's target velocity as provided by the high-level tactical planner and the velocity of the agent when passing a checkpoint in a single race. For the E2E agent, although it does not use a hierarchical structure, we still run an MCTS high-level tactical planner with an identical configuration in the background to simulate what the target lanes and velocities would be in its place and compare it to the actual lanes and velocities of the E2E agent's state at each checkpoint.

For every pair of controllers, we conduct 50 head-to-head races on each of the two tracks. Overall, every controller participates in 400 races. 
\subsection{Results}
\begin{figure*}
  \centering
  \includegraphics[width=0.9\textwidth]{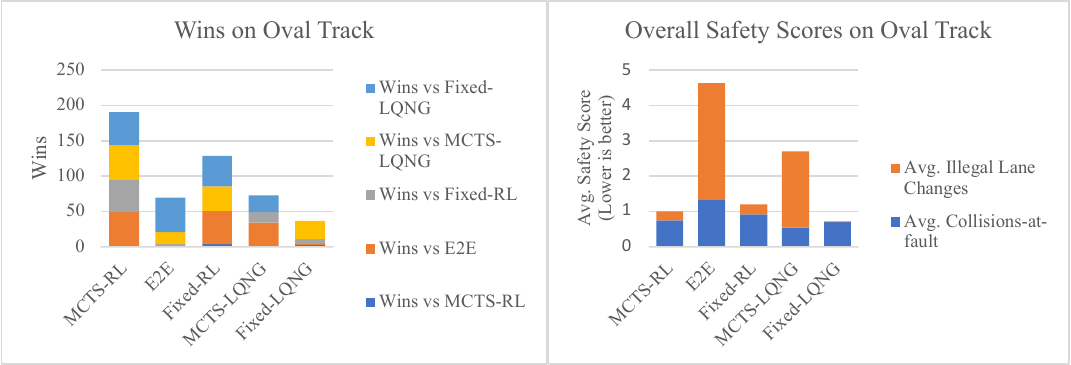}
  \caption{The graph on the left shows the number of wins for each agent type against each opposing agent type, and the graph on the right shows their safety scores when racing on the oval track.}
  \label{fig:results_oval}
  \bigskip
  \includegraphics[width=0.9\textwidth]{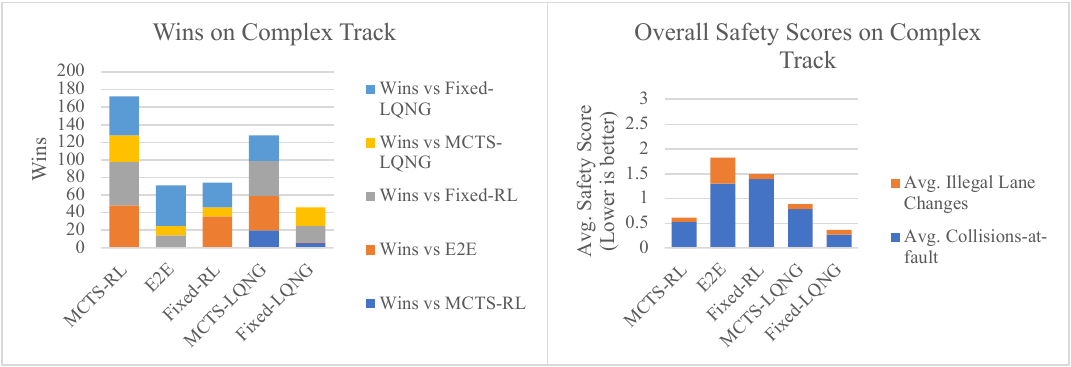}
  \caption{The graph on the left shows the number of wins for each agent type against each opposing agent type, and the graph on the right shows their safety scores when racing on the complex track.}
  \label{fig:results_complex}
\end{figure*}

Based on the plots in Figures \ref{fig:results_oval} and \ref{fig:results_complex} and Table \ref{tab:tldv_data}, we conclude the following key points. We also provide a supplemental video\footnote{\vidurl} demonstrating our agents in action. Detailed results used to produce Figures \ref{fig:results_oval} and \ref{fig:results_complex} are provided in the Appendix.

\begin{table}[!ht]
    \centering
    \begin{tabular}{|l|l|l|l|l|l|}
    \hline
        \textbf{Agent Type} & \textbf{$\mu_{\Delta_r}$} & \textbf{$\sigma_{\Delta_r}$} &\textbf{$\mu_{\Delta_b}$} & \textbf{$\sigma_{\Delta_s}$} & \textbf{Samples} \\ \hline
        MCTS-RL & $1.257^*$ & 0.256 & 2.08 & 0.307 & 389 \\ \hline
        E2E & $2.217^*$ & 0.239 & 2.02 & 0.325 & 391 \\ \hline
        MCTS-LQNG & $0.454^*$ & 0.101 & $0.64^*$ & 0.604 & 384 \\ \hline
    \end{tabular}
    \caption{The mean and standard deviations in the average target lane distance and average target velocity difference per checkpoint between the high-level planner's target lane and velocity and the true position and velocity at each checkpoint, respectively. The number of samples differs between each agent type as the agents did not finish every race. Results with an asterisk ($*$) indicate pair-wise statistical significance with all other agent types using a two-tailed t-test where the null hypothesis is that both agent types are equivalent and a p-value of 0.05.}
    \label{tab:tldv_data}
\end{table}

\begin{enumerate}[wide, labelindent=0pt]
\item \textbf{The proposed hierarchical variants outperformed their respective baselines.} 

The results amongst MCTS-RL, Fixed-RL, and E2E show the effectiveness of our proposed hierarchical structure. While all three of the MARL-based agents were only trained on the oval track, MCTS-RL was able to win the most head-to-head races while also maintaining the best safety score by better adapting its learning. Comparing the baselines against each other, Fixed-RL also has more wins and a better safety score than E2E across both tracks. This result indicates that formulating the game in a hierarchical structure is advantageous. It suggests that a straightforward task of trajectory tracking with collision avoidance is much easier to learn and generalize to unseen scenarios for a deep neural network. On the other hand, E2E was required to learn to execute tactical plans in addition to encoding the rules of the game. As a result, it likely experienced some level of overfitting to tactics that were successful only on the oval track and could not effectively apply the training to a never before seen environment.

Next, we compare MCTS-LQNG and Fixed-LQNG. Although MCTS-LQNG has a worse overall safety score, it has 25\% more wins when aggregated over both tracks. Fixed-LQNG has a similar number of overall wins on the oval track, but when the racetrack is more complicated, Fixed-LQNG quickly becomes inferior. The oval track has just one main racing line, but there exist many reasonable racing lines in the complex track that must be considered to be competitive. MCTS-LQNG accounts for these trajectories by using the high-level MCTS planner and is, therefore, more successful in its races against MARL-based agents on the complex track with four times the number of wins against them compared to the Fixed-LQNG agent. MCTS-LQNG considered trajectories that could result in overtakes when opponents made mistakes from any part of the track. On the other hand, Fixed-LQNG was forced to rely on opponents making mistakes that were not along its fixed trajectory to make overtakes. However, considering alternative lines also attributes to the main difference in their safety scores. Both have similar collision-at-fault scores, but MCTS-LQNG has more illegal lane changes.  

\item \textbf{Following the strategies proposed by the high-level tactical planner results in better performance.}

Studying the target lane distance and target velocity difference metrics in Table \ref{tab:tldv_data} in addition to the wins and safety scores in Figures \ref{fig:results_oval} and \ref{fig:results_complex}, we recognize that more closely following the high-level tactical plan results in more wins and better safety scores. Almost all of the average lane distance and target velocity differences were smaller for the MCTS-based agents, and they both have more wins and a better safety score than the E2E agent across both tracks. However, there seems to be a point of diminishing returns. Although MCTS-LQNG more closely follows the high-level strategy, it still has a worse safety score and win count than MCTS-RL. This result suggests that the high-level strategy is a good plan, but it is not necessarily always optimal and faithful with respect to the rules. This relates to the end of Section \ref{sec:ll_form} where we discuss how the high-level and low-level planners run independently and how stale information from the high-level planner might cause it to propose a strategy that might break the rules. These results find that the low-level planner must still bear some responsibility to reason strategically instead of blindly following the high-level plan. This conclusion and the comparison between the MARL and LQNG low-level planners are discussed in more detail in the following point.

\item \textbf{MARL is more successful and robust than LQNG as a low-level planner.}  

Overall, the MARL-based agents outperformed their LQNG-based counterparts in terms of both key metrics: wins and safety scores. However, this result is likely due to our simplifications involving a time-invariant linearization around the initial state of each agent, meaning the approximation is only valid for a very short time horizon. Therefore, the LQNG-based agents could only rely on braking/acceleration instead of yaw rate to avoid collisions. As a result, the weights in the objective of the LQNG formulation are set conservatively to emphasize avoiding collisions. This setup also implies that LQNG-based agents often concede in close battles and thereby lose races because of the high cost in the planning objective of driving near another player even if there is no collision. 

While Fixed-LQNG has a better safety score than Fixed-RL, MCTS-RL has a significantly better safety score than MCTS-LQNG. Just in terms of collision avoidance, both RL-based agents have worse numbers because the LQNG-based agents are tuned to be conservative. However, MCTS-LQNG has significantly increased illegal lane changes per race compared to MCTS-RL while Fixed-LQNG has slightly fewer illegal lane changes per race compared to Fixed-RL. As discussed previously, the fixed trajectory agents do not consider alternative racing lines, so they rarely break the lane-changing limit rule in the first place. In the MCTS-LQNG case, the high-level planner runs in parallel with the low-level and at a lower frequency. As a result, the calculated high-level plan uses slightly out-of-date information and does not account that the low-level controllers have already made choices that might contradict the initial steps in the plan. This mismatch causes the LQNG-based controller to break the lane-changing rules more frequently by swerving across the track to immediately follow the high-level plan when it is updated. MCTS-RL is more robust to this situation because they have those safety rules encoded in their reward structures, albeit with smaller weights. They do not track the waypoints exactly and learn to smooth the trajectory produced by the high-level plan and the live situation in the game.

\item \textbf{MCTS-RL outperforms all other implemented controllers.}  

 Aggregating the results from both tracks, MCTS-RL recorded a win rate of 90\% of the 400 head-to-head and the second-best safety score, only behind the conservatively tuned Fixed-LQNG agent. It combined the advantage of having a high-level planner that evaluates long-term plans and a low-level planner that is robust to the possibility that the high-level plans may be out of date. For example, Figure~ \ref{fig:mctsrl:overtake} demonstrates how the high-level planner provided a long-term strategy, guiding the agent to give up an advantage at present for a greater advantage in the future when overtaking. The RL-based low-level planner approximately follows the high-level strategy in case stochasticity of the MCTS algorithm yields a waypoint that seems out of place (e.g., the checkpoint between $t=3$ and $t=4$ in Figure~ \ref{fig:mctsrl:overtake}). Furthermore, MCTS-RL is also successful at executing defensive maneuvers as seen in Figure~ \ref{fig:mctsrl:defense} due to those same properties of long-term planning and low-level robustness. Both of these tactics resemble strategies of expert human drivers in real head-to-head racing.
 \end{enumerate}

 \begin{figure*}%
\centering
  \begin{tabular}{ c @{\hspace{20pt}} c }
    {\includegraphics[width=0.63\textwidth]{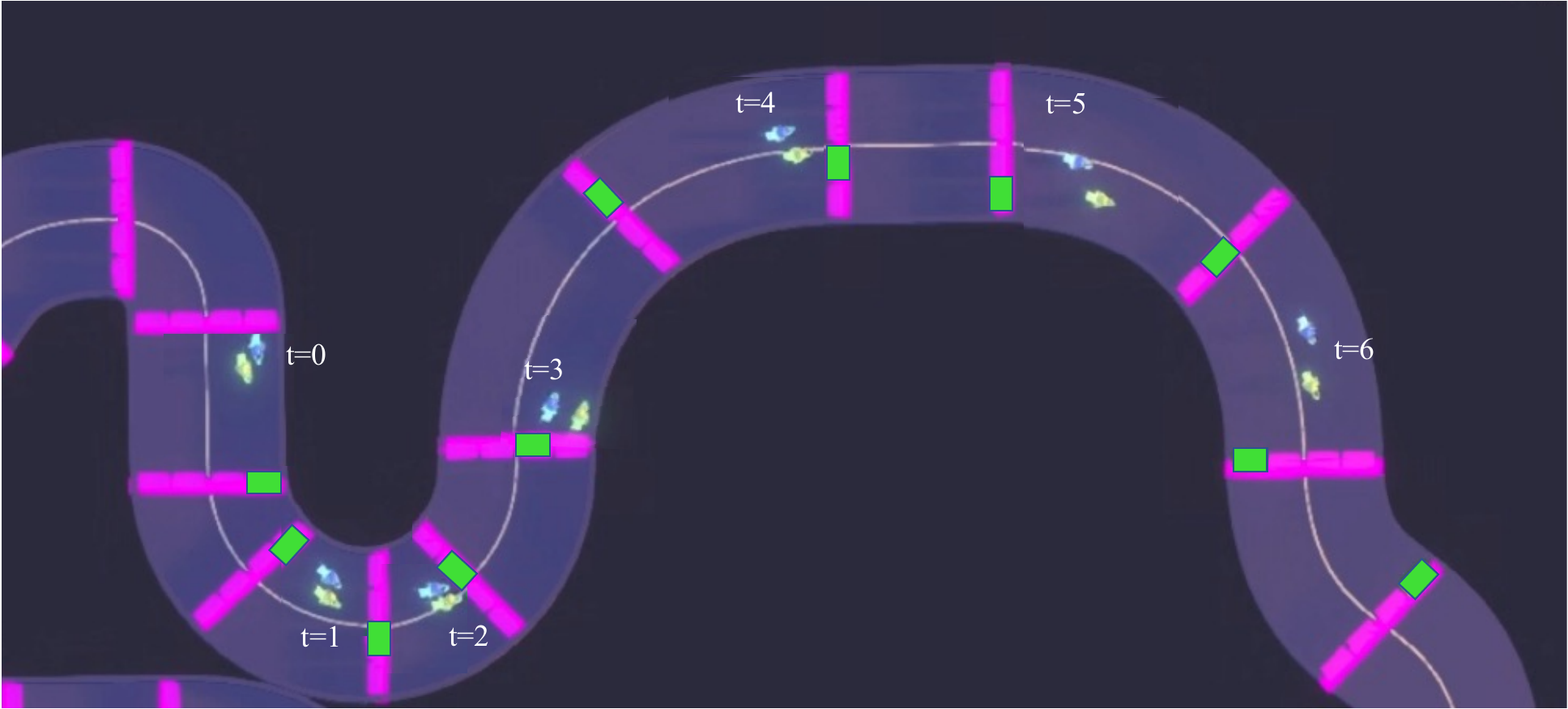}} &
    {\includegraphics[width=0.31\textwidth]{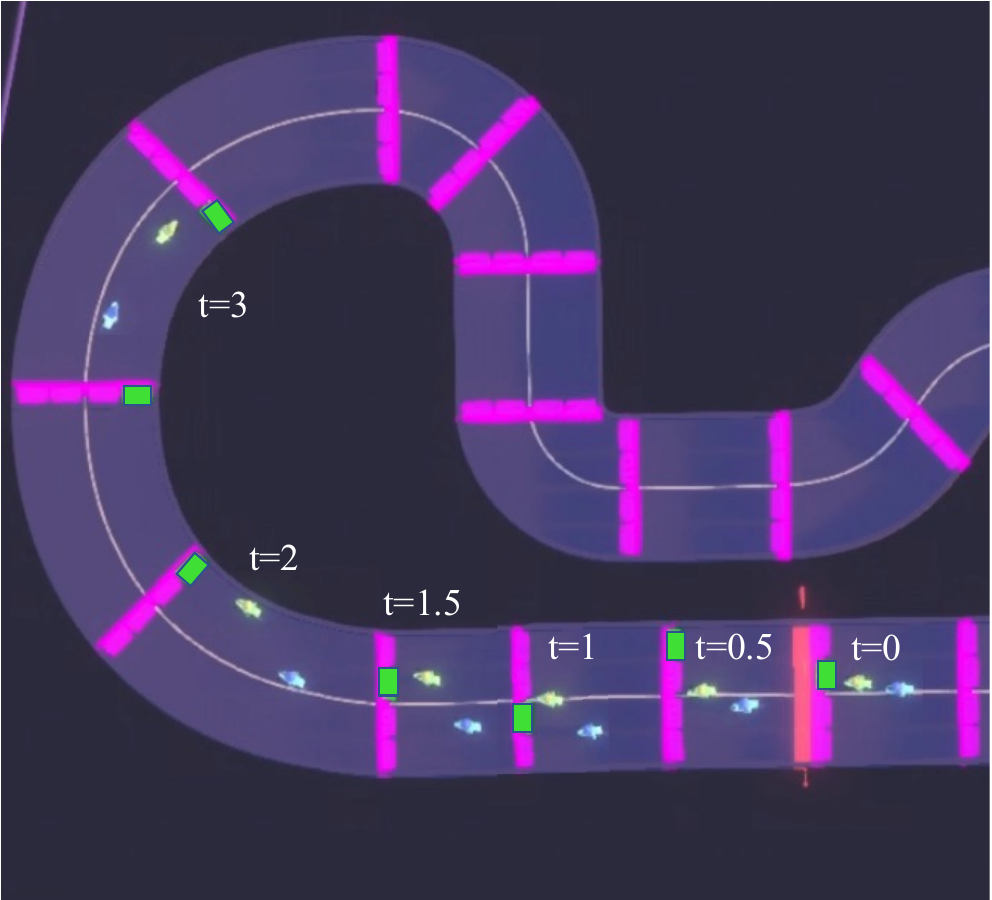}} 
    \\
    \small (a) &
      \small (b)
  \end{tabular}
    \caption{(a) An overtaking maneuver executed by the MCTS-RL agent (green) against the E2E agent (blue) on the complex track. Notice how, from $t=0$ to $t=2$, the MCTS-RL agent gives up a slight advantage and takes a wider racing line on the first turn. However, the exit of the wide racing line of the first turn places the MCTS-RL agent at the inside of the next two turns, where it gains an even greater advantage when passing the E2E agent from $t=3$ to $t=6$. The green boxes along each checkpoint also highlight the long-term plan calculated by the MCTS planner for this tactic. (b) A defensive maneuver executed by the MCTS-RL agent (green) against the E2E agent (blue) on the complex track.  Before reaching the turn, the MCTS planner chooses to switch one lane to the left first ($t=0$ to $t=1$) and then move to the right for the inside of the turn staying within the changing limit. This motion forces the E2E agent to make an evading move to avoid collision as it is behind the other agent thus responsible for collision avoidance. It must take an even wider turn, thus increasing the overall gap at the end. The green boxes along each checkpoint highlight the long-term plan calculated by the MCTS planner for this tactic.}%
    \label{fig:mctsrl}%
    \extralabel{fig:mctsrl:overtake}{a}
    \extralabel{fig:mctsrl:defense}{b}
\end{figure*}

\section{Conclusion}
We develop a hierarchical controller for head-to-head autonomous racing that adheres to safety and fairness rules found in real-life racing and outperforms other common control techniques such as purely optimization-based or purely learning-based control methods. Our high-level planner constructs long-term trajectories that abide by the introduced complex rules about collision avoidance and lane changes. As a result, we design an objective for the low-level controllers to focus on tracking the high-level plan, which is an easier problem to solve compared to the original racing-game formulation. Our hierarchical method outperforms baselines, representing common approaches to autonomous racing, both in terms of winning head-to-head races and a safety score measuring obedience to the rules of the game. Finally, we observed the controller exhibiting maneuvers resembling those performed by expert human drivers such as planning for delayed but more effective overtaking maneuvers and defensive blocking. We showed that game-theoretic hierarchical control is an effective method for path planning in autonomous racing.

Future work should introduce additional high-level and low-level planners and investigate policy-switching hierarchical controllers where we switch between various high and low-level controllers depending on the state of the game. Furthermore, the assumption of perfect self and opponent state information should be relaxed to further represent the reality of autonomous racing. In addition, a multi-agent scenario might also be studied with more than two players. One might also study the scalability of this approach beyond karts driving at \SI{15}{\meter\per\second} and consider high-speed scenarios at the limits of handling. Lastly, the concept of game-theoretic hierarchical control design can be extended to other multi-agent systems applications where there exist complex rules such as energy grid systems or air traffic control. Constructing a discrete high-level game allows for natural encoding of the complex constraints, often involving discrete components, to find an approximate solution that can warm start a more precise low-level planner.

\section{Acknowledgements}
This work was supported in part by National Science Foundation grants 230483 and 2211432; the Office of Naval Research grant N00014-23-1-2505; and the Army Research Laboratory grant W911NF-23-2-0011.

\bibliographystyle{apalike}
\bibliography{refs}

\section*{Appendix} \label{sec:app}
\subsection*{Multi-Agent Reinforcement Learning Controller Details}
We provide the details of the multi-agent reinforcement learning agents (MARL) in the following subsections. Note, that all of the agents using a MARL-based controller, E2E, MCTS-RL, and Fixed-RL, used identical reward structures, network architectures, and training procedures. The only differences in their setup were the weights selected for the various reward parameters to reflect the different objective functions of the underlying mathematical formulation of each type of agent. Recall that the MARL-based agents' observations include perfect state information for all players (including $(x, y)$ position, $v$ velocity, lane ID $a$, ``recent'' lane change count $e$, and last passed checkpoint $r$) and 9 LIDAR rays, whose distances we refer to as $I_1, \ldots, I_9$. Furthermore, we also assume players know the overall time elapsed in the game $t$, and the maximum time horizon $T$. 
\subsubsection*{Reward Structure Formulation}
We outline the specifics of the reward and penalty functions $R(\cdot)$ that use on one or more weight parameters denoted by $\omega_i$. Our rewards and penalties are categorized into two types:
\begin{enumerate}
\item For every time step in the environment, we provide the following rewards and penalties:
\begin{itemize}
    \item A reward for driving at higher speeds. The reward scales based on driving close to the top speed of the kart.
    \begin{equation*}
        R_{\text{speed}}(\omega_1) = \omega_1 \frac{v}{v_{\text{max}}}
    \end{equation*}
    \item A reward for moving towards the next checkpoint is denoted by $r*$. We use the three-dimensional velocity vector of the agent and take the dot product with the vector between the agent's position and the next checkpoint position.
    \begin{equation*}
        R_{\text{direction}}(\omega_1) = \omega_1 (\langle v_x, v_y \rangle \cdot \langle r^*_x-x , r^*_y - y\rangle)
    \end{equation*}
    \item A penalty for exceeding the lane changing limit. We use an indicator function to determine if the player is in the straight region of the track $\mathcal{S}$ and whether the lane changing limit $L$ is exceeded.
    \begin{equation*}
        R_{\text{swerve}}(\omega_1) = -\omega_1 \mathds{1}_{(x,y) \in \mathcal{S} \wedge e > L}
    \end{equation*}
    \item A penalty for being within $h$ meters of the wall. We use an indicator function $\mathds{1}_{I_j < h \wedge I_j \text{hit wall}}$ that determines if the LIDAR reading is below $h$ and if whether the LIDAR bounced off a player or a wall.
    \begin{equation*}
        R_{\text{wall-hit}}(\omega_1) = -\sum_{j=1}^9 \omega_1 \mathds{1}_{I_j < h \wedge I_j \text{hit wall}}
    \end{equation*}
    \item A penalty for being within $h$ meters of another player. Using a similar indicator function from above, if any LIDAR ray in that set hits another player within a distance $h$, then the original player is penalized for being in a collision. In addition, we assume we have a set $\Theta$, which contains the indices of the LIDAR rays that point towards the front of the kart. There is an additional penalty if the LIDAR rays come from the subset $\Theta$ as that indicates some form of rear-end collision where the player would be at fault.
    \begin{small}
    \begin{equation*}
        R_{\text{player-hit}}(\omega_1, \omega_2) = -\sum_{j=1}^9 (\omega_1 \mathds{1}_{I_j < h \wedge I_j \text{hit player}} + \omega_2 \mathds{1}_{j \in \Theta})
    \end{equation*}
    \end{small}
\end{itemize}

\item When a player passes a checkpoint with index $r'$, we provide the following rewards and penalties:
\begin{itemize}
    \item A reward to teach the policy to pass as many checkpoints as possible before other players. The reward is scaled based on the order in which the checkpoint is reached.
    \begin{equation*}
        R_{\text{checkpoint base}}(\omega_1) = 
        \begin{cases}
            \omega_1 & \text{if first} \\
            0.75\omega_1 & \text{if second} \\
        \end{cases}
    \end{equation*}
    \item A reward based on the remaining time in the game to incentivize minimizing the time between checkpoints. This reward is also added (with a different weight parameter) to a shared reward value used by the posthumous credit assignment algorithm to incentivize cooperative behavior. 
    \begin{equation*}
        R_{\text{checkpoint time}}(\omega_1) = \omega_1\frac{T-t}{T}
    \end{equation*}
    \item A reward for being closer to the target lane $a'$ and velocity $v'$ for the passed checkpoint.
    \begin{equation*}
        R_{\text{checkpoint target}}(\omega_1, \omega_2) = \frac{\omega_1}{1.3^{|a-a'|\sqrt{(a'_x-x)^2+(a'_y-y)^2}}}
         + \frac{\omega_2}{1.1^{|v-v'|}}
       \end{equation*}
    \item A penalty for driving in reverse. We use an indicator function to determine if checkpoint index $r'$ is less than or equal to $r$ implying the player passed either the same checkpoint or an earlier checkpoint.
    \begin{equation*}
        R_{\text{checkpoint reverse}}(\omega_1) = -\omega_1\mathds{1}_{r' \leq r}
    \end{equation*}
\end{itemize}
\end{enumerate}

\subsubsection*{Training and Network Structure}
Implementation and training of the MARL-based agents rely on a Unity library known as ML-Agents \cite{mlagents}. We use implementations of proximal policy optimization and self-play provided by the library to train and configure the neural networks of our agents. The training environment consists of 8 copies of two sizes of oval tracks. Within each set of tracks, half of the training assumed a clockwise race direction and the other half assumed a counter-clockwise direction. Using two sizes of tracks ensures that the agents learn to make both sharp and wide turns, and using the two race directions allows the agents to learn to make both left and right turns. However, the training is limited to just those track configurations to prevent overfitting and evaluate how the various controllers generalize to unknown environments such as the complex track. 

\begin{figure*}
  \centering
  \includegraphics[width=0.9\textwidth]{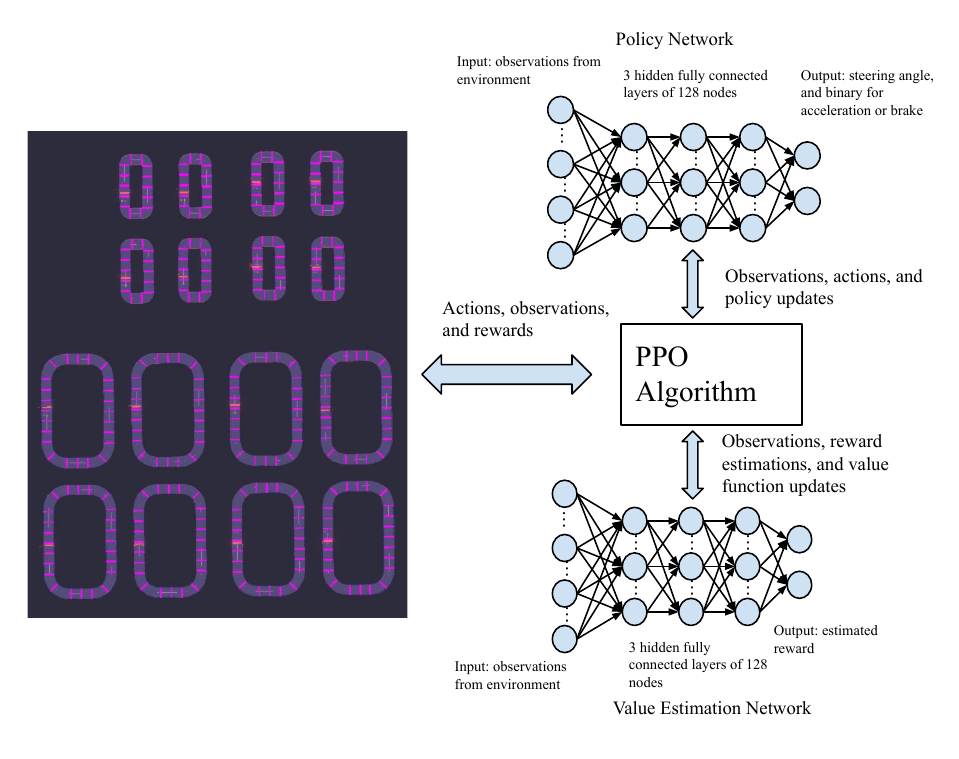}
  \caption{Visualization of the training environment and neural network structures used for all of the RL-based models. The only difference in the training of the three agents is the weights of the reward structures.}
  \label{fig:train_mod}
  \end{figure*}

\begin{sidewaysfigure}
\includegraphics[width=0.9\textwidth]{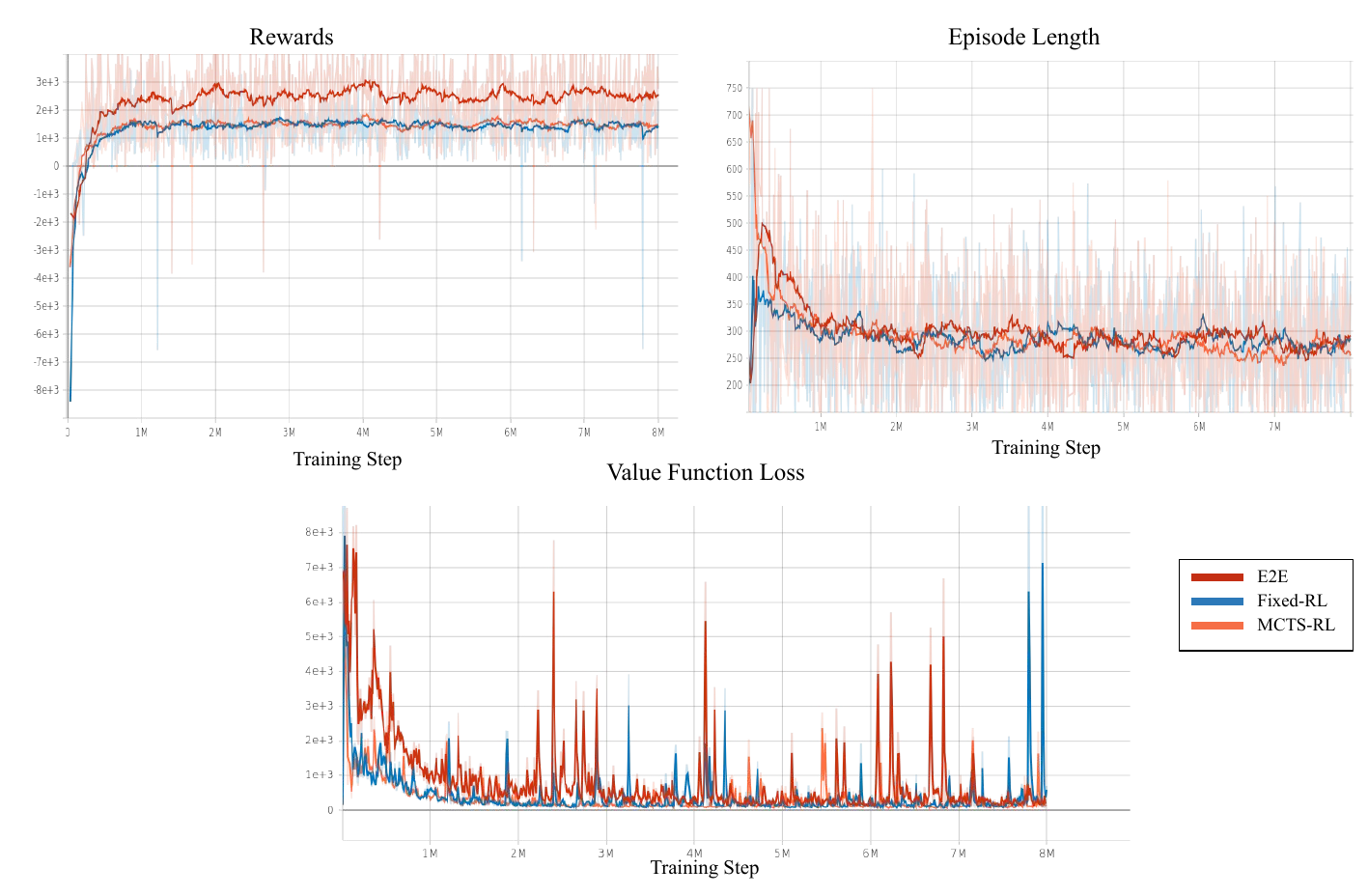}
  \caption{Plots of the rewards (top-left), episode lengths (top-right), and value function losses (bottom) convergence over the training of the RL-based models.}
  \label{fig:train_net}
\end{sidewaysfigure}

The agents share model inputs, policy and reward network sizes and structures, and model outputs. The input is a matrix consisting of stacked vectors of previously mentioned observations (own state, LIDAR rays, opponent state, checkpoint progress, etc.). Both the policy and reward networks consist of 3 hidden layers with 128 nodes each. Figure \ref{fig:train_mod} is a visualization of the described training environment, and Figure \ref{fig:train_net} presents the reward, episode length, and value function loss graphs across training showing their convergence. Note that the rewards scale varies amongst the three types of agents because the weights in the reward functions are different. However, all of the agents are trained to 8000000 steps and their rewards stabilized before reaching the step limit as seen in the graph. 

\subsection*{Detailed Tables of Results}
Below are tables with detailed head-to-head win results and safety metrics from the oval and complex track experiments. To determine the statistical significance of our data, we conduct pairwise two-tailed t-tests. For the head-to-head racing results, our null hypothesis is that the agents in each pair of agent types have an equal probability of winning a race. For the safety score results, our null hypothesis is that all agents have the same number of collisions-at-fault, number of illegal lane changes, and safety score per race. Statistically significant results are denoted using an asterisk ($*$) and evaluated using a p-value of 0.05.

\begin{table} [H]
    \centering
    \begin{tabular}{|l|p{1.5cm}|p{1.5cm}|p{1.5cm}|p{1.5cm}|p{1.5cm}|}
    \hline
        \textbf{Agent Type} & \textbf{Wins vs MCTS-RL} & \textbf{Wins vs E2E} & \textbf{Wins vs Fixed-RL} & \textbf{Wins vs MCTS-LQNG} & \textbf{Wins vs Fixed-LQNG} \\ \hline
        MCTS-RL & - & $49^*$ & $46^*$ & $49^*$ & $47^*$ \\ \hline
        E2E & $1^*$ & - & $3^*$ & $17^*$ & $49^*$ \\ \hline
        Fixed-RL & $4^*$ & $47^*$ & - & $35^*$ & $43^*$ \\ \hline
        MCTS-LQNG & $1^*$ & $33^*$ & $15^*$ & - & 24 \\ \hline
        Fixed-LQNG & $3^*$ & $1^*$ & $7^*$ & 26 & - \\ \hline
    \end{tabular}
    \caption{Oval track head-to-head wins. Results with an asterisk ($*$) indicate statistical significance using a p-value of 0.05.}
\end{table}

\begin{table} [!ht]
    \centering
    \begin{tabular}{|p{1.5cm}|p{2cm}|p{2cm}|p{1.5cm}|p{1.75cm}|p{1.25cm}|p{1.75cm}|p{0.9cm}|}
    \hline
        \textbf{Agent Type} & \textbf{Avg. Collisions-at-fault} & \textbf{Std. Dev. Collisions-at-fault} & \textbf{Avg. Illegal Lane Changes} & \textbf{Std. Dev. Illegal Lane Changes} & \textbf{Avg. Safety Score} & \textbf{Std. Dev. Safety Score} & \textbf{Races} \\ \hline
        MCTS-RL & $0.745$ & 1.3 & $0.255$ & 0.61 & $1$ & 1.49 & 200\\ \hline
        E2E & $1.33^*$ & 2.57 & $3.305^*$ & 2.22 & $4.635^*$ & 3.1 & 200\\ \hline
        Fixed-RL & $0.915$ & 1.26 & $0.285$ & 0.7 & $1.2$ & 1.6 & 200\\ \hline
        MCTS-LQNG & $0.54^*$ & 0.55 & $2.17^*$ & 1.7 & $2.71^*$ & 1.76 & 200\\ \hline
        Fixed-LQNG & $0.69$ & 0.46 & $0.03^*$ & 0.36 & $0.72^*$ & 0.55 & 200 \\ \hline
    \end{tabular}
    \caption{Oval track safety score statistics. Results with an asterisk ($*$) indicate pair-wise statistical significance with all other agent types using a p-value of 0.05.}
\end{table}

\begin{table}[!h]
    \centering
    \begin{tabular}{|l|p{1.5cm}|p{1.5cm}|p{1.5cm}|p{1.5cm}|p{1.5cm}|}
    \hline
        \textbf{Agent Type} & \textbf{Wins vs MCTS-RL} & \textbf{Wins vs E2E} & \textbf{Wins vs Fixed-RL} & \textbf{Wins vs MCTS-LQNG} & \textbf{Wins vs Fixed-LQNG} \\ \hline
        MCTS-RL & - & $48^*$ & $50^*$ & $30$ & $44^*$ \\ \hline
        E2E & $0^*$ & - & $14^*$ & $11^*$ & $46^*$ \\ \hline
        Fixed-RL & $0^*$ & $36^*$ & - & $10^*$ & $28$ \\ \hline
        MCTS-LQNG & $20$ & $39^*$ & $40^*$ & - & $29$ \\ \hline
        Fixed-LQNG & $6^*$ & $1^*$ & $18$ & $21$ & - \\ \hline
    \end{tabular}
    \caption{Complex track head-to-head wins. The total number of wins in a head-to-head match-up does not always add up to 50 because some races resulted in both agents failing to finish. Results with an asterisk ($*$) indicate statistical significance using a p-value of 0.05.}
\end{table}

\begin{table}[!ht]
    \centering
    \begin{tabular}{|p{1.5cm}|p{2cm}|p{2cm}|p{1.5cm}|p{1.75cm}|p{1.25cm}|p{1.75cm}|p{0.9cm}|}
    \hline
        \textbf{Agent Type} & \textbf{Avg. Collisions-at-fault} & \textbf{Std. Dev. Collisions-at-fault} & \textbf{Avg. Illegal Lane Changes} & \textbf{Std. Dev. Illegal Lane Changes} & \textbf{Avg. Safety Score} & \textbf{Std. Dev. Safety Score} & \textbf{Races} \\ \hline
        MCTS-RL & $0.525^*$ & 0.86 & $0.085$ & 0.328 & $0.61^*$ & 0.893 & 189 \\ \hline
        E2E & $1.3$ & 1.575 & $0.525^*$ & 0.98 & 1.825 & $1.848$ & 191 \\ \hline
        Fixed-RL & $1.39$ & 1.42 & $0.105$ & 0.494 & $1.495$ & 1.5 & 184 \\ \hline
        MCTS-LQNG & $0.785^*$ & 0.877 & $0.105$ & 0.352 & $0.89^*$ & 0.942 & 184 \\ \hline
        Fixed-LQNG & $0.275^*$ & 0.489 & $0.09$ & 0.335 & $0.365^*$ & 0.549 & 192 \\ \hline
    \end{tabular}
    \caption{Complex track safety score statistics. Results with an asterisk ($*$) indicate pair-wise statistical significance with all other agent types using a p-value of 0.05.}
\end{table}

\begin{table}[!ht]
    \centering
    \begin{tabular}{|l|p{1.5cm}|p{1.5cm}|p{1.5cm}|p{1.5cm}|p{1.5cm}|}
    \hline
        \textbf{Agent Type} & \textbf{Wins vs MCTS-RL} & \textbf{Wins vs E2E} & \textbf{Wins vs Fixed-RL} & \textbf{Wins vs MCTS-LQNG} & \textbf{Wins vs Fixed-LQNG} \\ \hline
        MCTS-RL & - & $97^*$ & $96^*$ & $79^*$ & $91^*$ \\ \hline
        E2E & $1^*$ & - & $17^*$ & $28^*$ & $95^*$ \\ \hline
        Fixed-RL & $4^*$ & $83^*$ & - & $45$ & $71^*$ \\ \hline
        MCTS-LQNG & $21$ & $55^*$ & $40$ & - & $53$ \\ \hline
        Fixed-LQNG & $9^*$ & $25^*$ & $18^*$ & $47$ & - \\ \hline
    \end{tabular}
    \caption{Cumulative head-to-head wins. The total number of wins in a head-to-head match-up does not always add up to 100 because some races resulted in both agents failing to finish. Results with an asterisk ($*$) indicate statistical significance using a p-value of 0.05.}
\end{table}

\begin{table}[!ht]
    \centering
    \begin{tabular}{|p{1.5cm}|p{2cm}|p{2cm}|p{1.5cm}|p{1.75cm}|p{1.25cm}|p{1.75cm}|p{0.9cm}|}
    \hline
        \textbf{Agent Type} & \textbf{Avg. Collisions-at-fault} & \textbf{Std. Dev. Collisions-at-fault} & \textbf{Avg. Illegal Lane Changes} & \textbf{Std. Dev. Illegal Lane Changes} & \textbf{Avg. Safety Score} & \textbf{Std. Dev. Safety Score} & \textbf{Races} \\ \hline
        MCTS-RL & $0.635$ & 1.11 & $0.17$ & 0.496 & $0.805^*$ & 1.24 & 389 \\ \hline
        E2E & $1.32$ & 2.13 & $1.92^*$ & 2.11 & $3.23^*$ & 2.91 & 391 \\ \hline
        Fixed-RL & $1.15$ & 1.37 & $0.195$ & 0.61 & $1.35^*$ & 1.56 & 384 \\ \hline
        MCTS-LQNG & $0.663$ & 0.741 & $1.14^*$ & 1.6 & $1.8^*$ & 1.68 & 384 \\ \hline
        Fixed-LQNG & $0.483^*$ & 0.519 & $0.06^*$ & 0.384 & $0.542^*$ & 0.577 & 392 \\ \hline
    \end{tabular}
    \caption{Cumulative safety score statistics. Results with an asterisk ($*$) indicate pair-wise statistical significance with all other agent types using a p-value of 0.05.}
\end{table}
\FloatBarrier
\end{document}